\documentclass{ws-procs975x65}            

\def\gsim{\mathrel{\rlap{\lower 4pt \hbox{\hskip 1pt $\sim$}}\raise 1pt
\hbox {$>$}}}
\def\lsim{\mathrel{\rlap{\lower 4pt \hbox{\hskip 1pt $\sim$}}\raise 1pt
\hbox {$<$}}}

\begin{document}
\title{Progenitors of type Ia supernovae}

\author{Keiichi Maeda}

\address{Department of Astronomy, Kyoto University, \\
Kitashirakawa-Oiwake-cho, Sakyo-ku, Kyoto 606-8502, Japan \\
E-mail: keiichi.maeda@kusastro.kyoto-u.ac.jp}

\author{Yukikatsu Terada}

\address{Graduate School of Science and Engineering, \\
Saitama University, 255 Shimo-Ohkubo, Sakura, Saitama 338-8570,Japan}

\begin{abstract}
Natures of progenitors of type Ia Supernovae (SNe Ia) have not yet been clarified. There has been long and intensive discussion on whether the so-called single degenerate (SD) scenario or the double degenerate (DD) scenario, or anything else, could explain a major population of SNe Ia, but the conclusion has not yet been reached. With rapidly increasing observational data and new theoretical ideas, the field of studying the SN Ia progenitors has been quickly developing, and various new insights have been obtained in recent years. This article aims at providing a summary of the current situation regarding the SN Ia progenitors, both in theory and observations. It seems difficult to explain the emerging diversity seen in observations of SNe Ia by a single population, and we emphasize that it is important to clarify links between different progenitor scenarios and different sub-classes of SNe Ia. 
\end{abstract}


\bodymatter

\section{White Dwarfs and Type Ia Supernovae}\label{aba:s1}

There is almost no doubt that Type Ia Supernovae (SNe Ia) are thermonuclear explosions of a C+O white Dwarf (WD) in a binary system. SNe Ia are one of the most matured standardized candles that led to the discovery of the accelerating expansion of the Universe,\cite{phillips1999,permutter1999,riess1998} and further improvement as the cosmological distance indicator, to the level to constrain the `Dark Energy Equation of State' by SN Ia observational data, requires the better understanding of the progenitor system and the explosion mechanism.  SNe Ia also play a key role in the chemical enrichment of galaxies and the Universe,\cite{matteucci1986,kobayashi1998} and our understanding of SNe Ia is directly linked to the origin of major chemical species. Involving the binary evolution as a major ingredient to form an immediate progenitor star, studying SNe Ia also provides a key input to clarify many unresolved issues in the binary evolution.\cite{ruiter2009}

Several progenitor scenarios have been proposed for SNe Ia. There are many variants in details, but they are basically categorized into the following classes; the single-degenerate (SD), double-degenerate (DD), and core-degenerate (CD) scenarios. In the SD scenario, a C+O WD  accretes materials from its non-degenerate companion star (either a main-sequence, MS, or a red giant, RG) and reach to the Chandrasekhar limiting mass.\cite{whelan1973,nomoto1982} The DD scenario involves a merger of two WDs,\cite{iben1984,webbink1984} and the CD scenario involves a merger of a WD and an asymptotic giant branch (AGB) star in a binary system.\cite{sparks1974,soker2015} 

\section{Diversity of SNe Ia}

\begin{figure}[t]%
\begin{center}
\parbox{4.8in}{\includegraphics[width=4.5in]{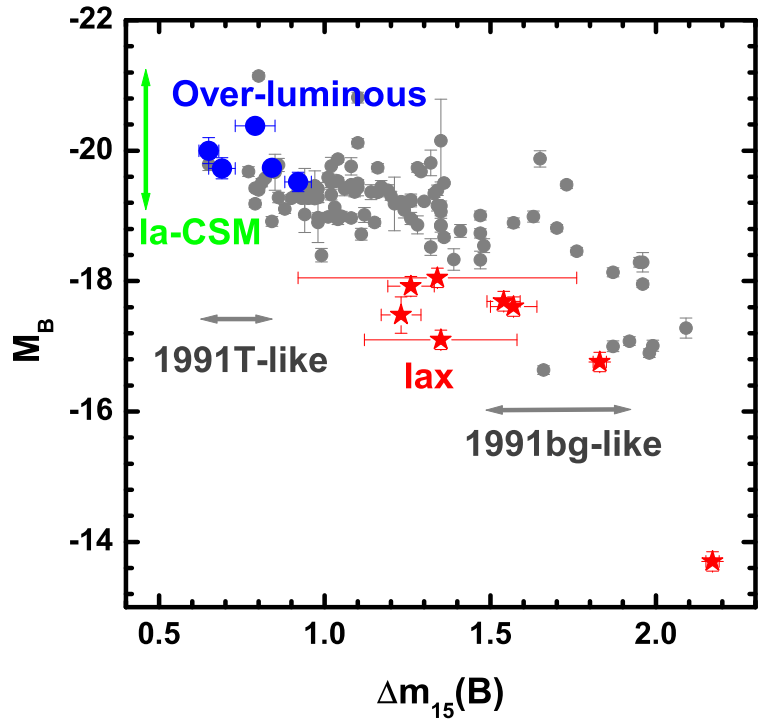}}
 \caption{Diversities of SNe Ia. The figure shows the declining rate ($\Delta m_{15} (B)$: the magnitude decline in the $B$-band in 15 days since the maximum light) and the peak absolute magnitude in the $B$-band ($M_{B}$) for a number of SNe Ia. The gray points are from the CfA sample\cite{hicken2009} which includes mostly normal SNe Ia but also some outliers (Note that the distances for these SNe in this figure are simply computed as the Hubble flow distances without taking into account any peculiar motion). Gray arrows indicate the ranges of the declining rates typically found for SN 1991T-like and 1991bg-like SNe Ia. Over-luminous SNe Ia (beyond SN 1991T-like) are indicated by blue points.\cite{howell2006,hicken2007,yamanaka2009,scalzo2010,chakradhari2014} SNe Iax are indicated by red stars.\cite{foley2013} A green arrow indicates the luminosity range for SNe Ia-CSM.\cite{silverman2013a}}
\label{f1}
\end{center}
\end{figure}

A classical question about the progenitor system of SNe Ia has been to identify a `single' correct path to SNe Ia. However, rapidly increasing observational data have been uncovering a vast diversity in natures of SNe Ia, and now a possibility that SNe Ia are indeed a mixture of the end products of different evolutionary paths has been raised by many researchers. A main issue then is to clarify a relation between different progenitor paths and different (observational) classes of SNe Ia. 

SNe Ia are relatively homogeneous objects. A scatter in the peak luminosity is mostly within an order of magnitude, or even much smaller if we focus on `normal' SNe Ia (Fig. 1). Spectra show some diversity, but basically they are well described by P-Cygni lines from intermediate and Fe-peak elements superimposed on a thermal continuum (in the early phase) or forbidden lines from Fe-peak elements with virtually no continuum (in the late, nebular phase). Furthermore, typical SNe Ia, or `normal' SNe Ia, show a strong correlation between the light curve declining rate and the peak luminosity (the so-called Phillips relation),\cite{phillips1999} and after correcting for this correlation their peak luminosities are quite uniform. This correlation is further extended to spectral features, where the appearance and strength of different lines can be put into the temperature sequence.\cite{nugent1995}

However, the recent development in the observational study has been leading to the discovery of outliers. Indeed, these SNe might not be intrinsically so rare. Figure 1 shows these outliers as compared to normal SNe Ia in the relation between the declining rate and the peak magnitude. The `classical' outliers are SN 1991T-like SNe on the bright end\citep{filippenko1992a} and SN 1991bg-like SNe in the faint end.\cite{filippenko1992b} Further `new classes' have been added into the classification scheme. Over-luminous SNe (sometimes called Super-Chandrasekhar SNe Ia) are so bright at the peak that the estimated mass of $^{56}$Ni synthesized in these events exceeds $\sim 1 M_{\odot}$ or even more, and the reasonable estimate of the total mass is beyond the Chandrasekhar mass.\cite{howell2006,hicken2007,yamanaka2009,scalzo2010,chakradhari2014} They typically show slow expansion velocities in the absorption lines while there are exceptions like SN 2006gz, and frequently a strong carbon feature is identified. 

Another class of the variants, which has been intensively studied in the last decade, is Type Iax SNe, originally classified as SN 2002cx-like SNe.\cite{li2003,phillips2007,sahu2008,foley2009} Their early-phase spectra are dominated by Fe lines at higher ionization state and characterized by bluer continuum than normal SNe Ia. The line velocities seen in SNe Iax are much lower than in normal SNe Ia. SNe Iax are typically faint, with the luminosities spanning in a wide range between $\sim -14$ and $\sim -18$ mag (Fig. 1: Note the point near the bottom-right edge of the figure). They do not follow the Phillips relation, but SNe Iax themselves may form a well-defined relation between the peak luminosity and the declining rate.\cite{foley2013}

SNe showing a signature of dense circumstellar materials (CSM) provide a link to the progenitor evolution. `Ia-CSM' is a sub-class of SNe IIn that show blue continuum and strong emission lines of Balmer series, which are clear signatures of SN ejecta expanding into dense CSM.\cite{hamuy2003,aldering2006,taddia2012,silverman2013a} Unlike SNe IIn from massive star explosions, the Ia-CSM class objects are explained as SNe Ia exploding within dense CSM. Interestingly, the SNe embedded in the Ia-CSM class are almost exclusively identified as SN 1991T-like SNe Ia.\cite{leloudas2015} 

\section{Nature of White Dwarfs in Binary Systems}
Before going deep into different progenitor scenarios, it is probably instructive to briefly summarize properties of observed Galactic WDs. Especially, the mass of a primary WD is of particular importance, since a mode of the explosion is sensitive to the mass of a primary star, together with the accretion rate (in the SD) and the mass of a secondary WD (in the DD). 

Intermediate polar-type (IP) cataclysmic variables (CV) are systems consisting of a strongly-magnetized WD as a primary and a main-sequence (MS) companion star. The companion star is typically not a massive MS, e.g., a M-type star with $\sim 0.1 M_{\odot}$ for the case of EX Hya, thus they are not typically considered to be a progenitor system of SNe Ia. There is, however, indeed suggestion that a massive C+O WD/M-type MS system could be a progenitor of SNe Ia for a system with a companion star having $> 0.4 M_\odot$.\cite{wheeler2012} Even for the systems whose total mass is not massive enough to lead to an SN Ia, studying their natures is highly interesting since the physical and evolutionary processes involved in these systems have many characteristic features/links shared by other systems like recurrent novae. 

In the IP systems, materials spilling over the Roche lobe of the MS companion are captured by the magnetic field of the WD and form hot plasma on the magnetic poles of the WD, emitting strong hard X-rays via thermal bremsstrahlung. The hot plasma in the accretion column has a multi-temperature structure,\cite{aizu1973,hoshi1973} as has been observationally confirmed.\cite{ezuka1999,terada2001} Since the plasma temperature directly reflects the depth of gravitational potential of the WD, an accurate estimate is possible through the X-ray spectral analyses on the gravitational potential, thus the ratio between the WD mass and radius.\cite{ishida1997} The mass and radius can be further separately obtained through the detailed accretion column model taking into account the multi-temperature structure.\cite{yuasa2010}. Spectra of two of the most famous IPs, EX Hya and V1223 Sgr observed with {\em Suzaku} can be reproduced by an IP X-ray spectral model, as a combination of a thermal component and a reflection component.\cite{hayashi2014a,hayashi2014b} For EX Hya and V1223 Sgr, the WD masses have been measured to be $0.65^{+0.11}_{-0.12}\,M_{\odot}$ and $0.91^{+0.08}_{-0.03}\,M_{\odot}$, respectively. 

Novae are one of the main ingredients in the SD scenario (see \S 4 for details). Among them, recurrent novae have been proposed to be one possible immediate progenitor system to SNe Ia -- the short time scale between successive nuclear runaways indicates a massive nature of a primary WD, as highlighted by the derived primary WD masses of $\sim 1.35 - 1.38 M_{\odot}$ for several recurrent nova systems obtained through light curve models.\cite{hachisu2000,hachisu2001} In the nova path to an SN Ia in the SD scenario, another key requirement is that the net accretion rate (i.e., the mass of accreted materials during the quiescent phase minus that ejected at a nova eruption) is positive. The general answer for this is not yet clear,\cite{schaefer2011} but given the nearly Chandrasekhar mass WD as derived for some recurrent nova systems, at least a part of systems undergoing novae are viable candidates as an SN Ia progenitor. 

WD-WD binaries are also important progenitor candidates for SNe Ia (in the DD scenario). Recent study on one possible WD-WD binary system, Hen 2-428, provided breakthrough in this respect.\cite{santander2015} The combination of the photometric and radial velocity monitoring of Hen 2-428, a binary system with an orbital period of 4.2 hours hosting a bipolar planetary nebula (PN), has allowed to derive the system's orbital parameters. The two post-AGB binary members have similar characteristics, with the mass of each derived to be 0.88$\pm$0.13 M$_\odot$. The stars will merge in $\sim$700,000 year. This makes Hen 2-428 the first promising double-degenerate system with a total mass clearly above the Chandrasekhar limit.

\section{Progenitor Systems and Explosion Mechanisms of SNe Ia}

\begin{figure}[t]%
\begin{center}
\parbox{4.8in}{\includegraphics[width=4.5in]{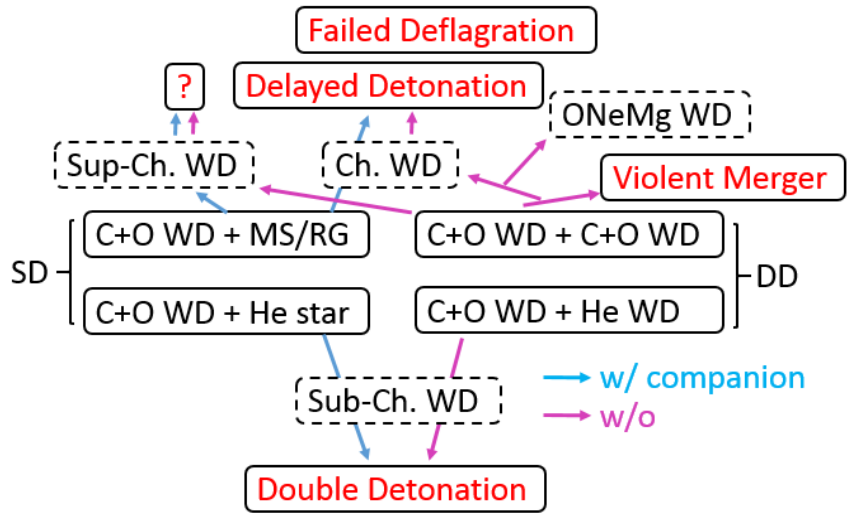}}
 \caption{A schematic `flowchart' of SN Ia progenitor models related to the SD and DD scenarios. See the text for details.}
\label{f2}
\end{center}
\end{figure}

Proposed evolutionary models for SN Ia progenitors are roughly divided into two categories: the SD\cite{whelan1973,nomoto1982} and DD.\cite{iben1984,webbink1984} In this section, we summarize basic ideas and related model variants. As demonstration, Figure 2 shows a (simplified) flowchart linking the progenitor scenarios, expected natures of the immediate progenitor WD, and the modes of the explosion. Figures 3 and 4 illustrate some explosion models related to the SD scenario (Fig. 3) and the DD scenario (Fig. 4), respectively. The so-called CD scenario fits into neither of the two scenarios, and ample discussion on the expected consequences and the possible observational constraints/indications for the scenario are found elsewhere.\cite{sparks1974,soker2015}

\subsection{Single degenerate scenario}

The most popular branch of the SD scenario involves a system of a primary C+O WD with either a MS ($\sim 2 M_{\odot}$) or RG ($\sim 1 M_{\odot}$).\cite{li1997,hachisu1999a,hachisu1999b} If the mass of the primary WD can increase through Roche lobe accretion and reach to the Chandrasekhar limiting mass, the primary star could explode as an SN Ia by the carbon burning near the center of the WD triggering the nuclear runaway.\cite{whelan1973,nomoto1982,nomoto1984}

A key question for these systems to be a viable SN Ia progenitor is whether the primary WD can reach to the Chandrasekhar limiting mass, and the accretion rate is a key. It has been shown that for a certain range of the accretion rate ($\sim 10^{-7} M_{\odot}$ yr$^{-1}$), the WD mass can grow thanks to stable hydrogen burning on the WD surface. Below this rate, the WD will experience explosive burning leading to novae. Beyond this rate, the WD was expected to experience a common-envelope (CE) phase with its companion star and would not lead to an SN Ia explosion. This `fine-tuning' in the accretion rate had been regarded to be one of the key difficulties in the SD model. However, it has been proposed that the high accretion rate beyond the classical stable burning limit would not lead to the CE phase, once the effect of a WD wind is taken into account.\cite{hachisu1996} The `nova' path could also lead to an SN Ia explosion, given the observational Galactic counterparts like U SCO 2010. 

Once the C+O WD reaches nearly the Chandrasekhar mass, it is likely that thermonuclear runaway starts near the center of the WD.\cite{nomoto1984} Theoretically it is not clear if the thermonuclear flame proceeds as a deflagration or detonation wave from the first principle. However, an immediate detonation deep in the Chandrasekhar-mass WD produces too much Fe-peak elements (including $^{56}$Ni which decays to Fe, powering the SN emission) to be compatible to the luminosities of individual SNe Ia and the Galactic chemical evolution.\cite{arnett1969,woosley1986} Thus, the deflagration is left as a possibility. However, the deflagration alone is thought to be too weak,\cite{timmes1992,livne1993} unable to produce a sufficient amount of $^{56}$Ni. These considerations lead to the `delayed detonation model',\cite{khokhlov1991,arnett1994,hoeflich1995,hoeflich1998,iwamoto1999} in which the deflagration is turned into detonation at the fuel density of $\sim 10^7$ g cm$^{-3}$. While these scenarios have been intensively studied in 1D, the mode indeed involves a multi-dimensional nature. Thermonuclear sparks are ignited within a convective core. The WD convective core might well have a dipole mode,\cite{woosley2004,kuhlen2006} and thus the distribution of the sparks is not necessarily represented by a spherical symmetry.\cite{roepke2007b} Once the deflagration flame develops, this should induce turbulent motion (e.g., see Ref.~\refcite{roepke2007a} for a review). The delayed detonation model has thus been studied with 2D and 3D simulations. \cite{kasen2009,maeda2010a,seitenzahl2013}

\begin{figure}[t]%
\begin{center}
 \parbox{4.8in}{\includegraphics[width=4.5in]{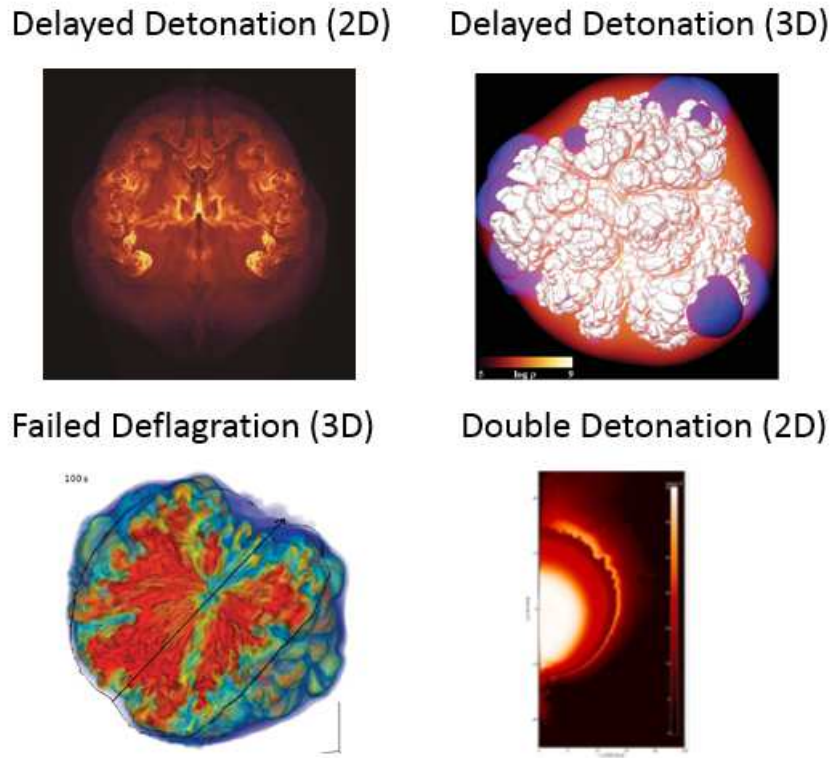}}
 \caption{A gallery showing some examples of explosion models (mostly) related to the SD scenario; a 2D delayed detonation model,\cite{maeda2010a} 3D delayed detonation model (reproduced by permission of the Oxford Journals),\cite{seitenzahl2013} 3D failed deflagration model (reproduced by permission of the Oxford Journals),\cite{kromer2013} and a 2D double detonation model (reproduced by permission from Astronomy \& Astrophysics, ESO).\cite{fink2010} Indeed, variants of the DD model would also lead to similar (or same) configurations (see the text for details). 
}
\label{f3_km}
\end{center}
\end{figure}

Recently, an interesting possibility is raised about whether the deflagration is turned into the detonation. If this would not take place, the deflagration would not produce enough nuclear energy to unbind the whole WD. This would eject a part of the WD but leave the rest of the WD as a compact remnant.\cite{kromer2013} This so-called `failed deflagration model' has been gaining attention as a model of SNe Iax. 

Another variant to the SD model exists for a case of a (non-degenerate) helium donor. Helium accumulated on the WD surface could induce unstable triple-alpha reactions leading to detonation. If the detonation is sufficiently strong, it would lead to the disruption of the whole WD.\cite{livne1990,woosley1994,shen2009} In 1D simulations, it could happen when the (spherically symmetric) shock wave penetrating deep into the WD converges at the WD center, and if the temperature at the convergence is sufficiently high to trigger the carbon detonation. If the distribution of the He detonating spots far deviates from spherical symmetry, it is more likely that the convergence of the shock waves takes place on the opposite side of the WD surface and then the converged shock further penetrates into the WD interior to trigger the carbon detonation, as shown by 2D simulations.\cite{fink2010} Irrespective of the details, this is called the double-detonation scenario. This scenario does not require a Chandrasekhar mass WD, and it is classified as a `sub-Chandrasekhar scenario'. 

The angular momentum can have an important role in the SD systems, but just recently its effect has been considered seriously. The accretion should keep the WD spinning at a high rate. Thus, the WD could indeed grow beyond the Chandrasekhar mass (the `spin-up/down scenario').\cite{yoon2005,hachisu2012,boshkayev2013,wang2014} After the accretion rate decreases, the WD could lose the angular momentum, e.g., by the magnetic breaking. Then the density and temperature in the WD interior would increase, to the point to ignite the carbon burning runaway. In this scenario, the immediate progenitor is a super-Chandrasekhar mass WD, probably the mass exceeding the Chandrasekhar mass of a non-rotating WD by a few percent. If this happens, an RG companion likely evolves to a He WD before the explosion of the primary WD.\cite{benvenuto2015}

\subsection{Double-degenerate scenario}

A main ingredient of the DD scenario is a close binary of double WDs that merge within the Hubble time through the gravitational wave (GW) emission. The classical DD model requires a system in which the total mass of the binary members exceeds the Chandrasekhar mass.\cite{iben1984,webbink1984} In this picture, the merger product is represented by a massive WD (initially the primary) accreting materials from an envelope and accretion disk/torus (initially the secondary but disrupted at the merger). The system then resembles (the Chandrasekhar mass WD model of) the SD model, but with a significant difference in the high accretion rate for the DD system. If the primary WD can evolve to the Chandrasekhar mass WD, the system would explode as an SN Ia through the delayed-detonation mode. 

Theoretically, whether this path is viable or not is unclear. Especially, it has been suggested, and shown in one-dimensional simulations, that the high accretion rate onto the WD surface would lead to the off-center carbon deflagration that converts the whole WD into an ONeMg WD.\cite{saio1985} Since the ONeMg WD, when reaching nearly to the Chandrasekhar mass, will lead to an electron capture-induced collapse to form a neutron star,\cite{nomoto1991,dessart2006} it will not explode as an SN Ia. However, at least for a certain parameter space the system could avoid the off-center carbon deflagration and evolve into a C+O Chandrasekhar WD, as shown by some 2D simulations.\cite{yoon2007} 

An alternative path to an SN Ia under the DD systems is the so-called violent merger scenario.\cite{pakmor2010,pakmor2012} At the merger, an accretion stream can dynamically produce a high temperature in hitting the primary WD surface. If the temperature is so high to induce the carbon detonation, the whole system would explode by carbon burning during the merger process. The temperature at the `hot spot' is strongly dependent on the binary parameters, and this scenario is only efficient for the systems where the total mass is indeed much beyond the Chandrasekhar mass.\cite{sato2015} This is indeed one difficulty in this scenario to explain the main population of SNe Ia, given the expected low frequency of such events.\cite{ruiter2011,ablimit2016} It should be emphasized that the model is indeed the `sub-Chandrasekhar model' in terms of the density of the WD(s) on which the thermonuclear flames propagate. 

\begin{figure}[t]%
\begin{center}
 \parbox{4.8in}{\includegraphics[width=4.5in]{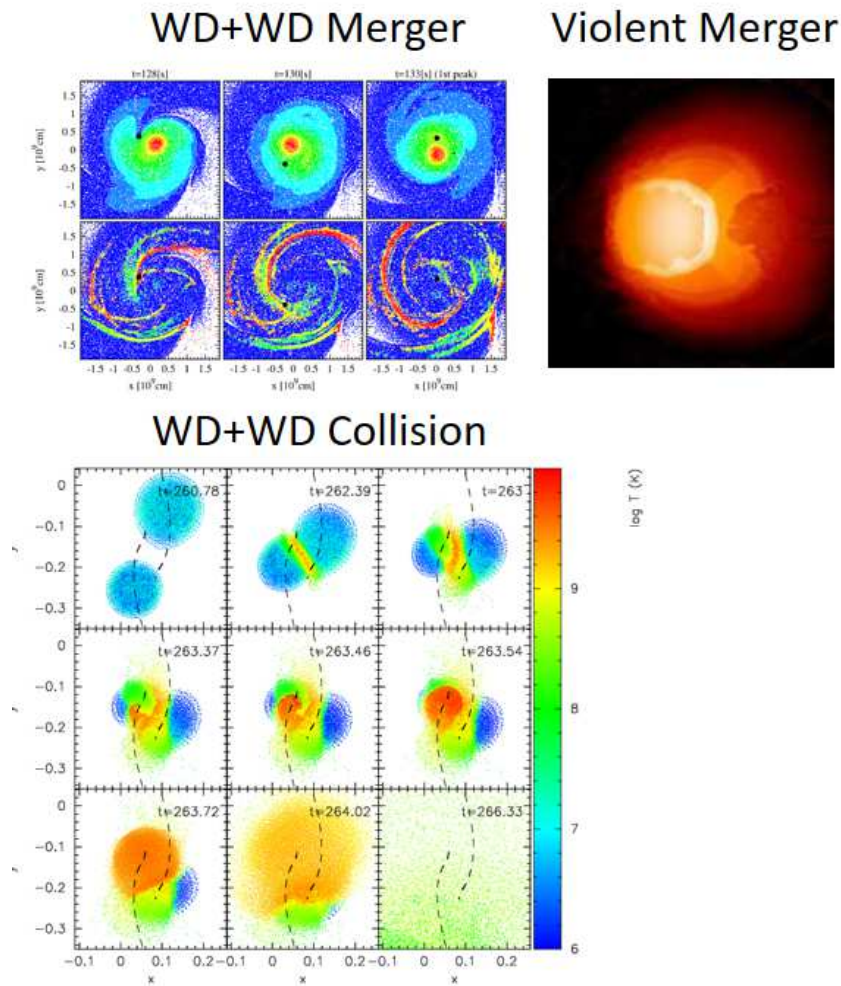}}
 \caption{A gallery for numerical examples related to the DD scenario. A binary merger of two C+O WDs following the orbital shrinking by a gravitational wave emission (left panel)\cite{tanikawa2015} may lead to various configurations, i.e., a single massive WD and a massive envelope and accretion disk (`classical DD scenario') or a prompt detonation following the merger (`violent merger' as shown in the right panel; reproduced by permission of the AAS).\cite{roepke2012} The bottom panel shows an example of a nearly head-on collision of two WDs as a result of dynamical interaction in a dense stellar environment (reproduced by permission of the Oxford Journals).\cite{aznar2013} 
}
\label{f4_km}
\end{center}
\end{figure}

A variant for the above scenario can be considered. In a triplet system that contains at least two WDs in a close orbit, the three body interaction could increase the ellipticity of the binary WD orbit. This would finally lead to nearly a head-on collision of two WDs.\cite{kushnir2013} The compression due to the impact would raise the temperature to the point where the carbon detonation is triggered. A similar situation is also realized in the high density environment like a globular cluster, where frequent WD-WD collisions would be expected.\cite{aznar2013} An example of this scenario is shown in Fig. 4, where the dynamical interaction of two WDs results in the disruption of both stars (for the WD masses of $0.8 M_\odot$ and  $1.0 M_\odot$).  

The above scenarios (both for the merger following the GW emission and for the head-on collision) consider a merger of two C+O WDs, but helium content can change the situation. If the secondary WD contains a larger amount of He on the surface, i.e., either a He WD or a C+O WD with a He envelope, the He detonation would be induced at the primary's surface. The evolution afterword is likely an analog to the double-detonation model in the SD scenario.

\section{Observational Constraints}

In this section, we summarize the state-of-the-art situation of observational constraints on the different scenarios for individual SNe (but note that the list could be biased by the authors' viewpoint). The constraints on the progenitor systems of individual SNe are largely divided into two categories -- one addressing a companion star and the other addressing the CSM environment. Finally, studying the electromagnetic features of individual SNe provides various constraints, especially sensitive to the mode of the explosion. A summary of the section is given in Table 1 (for an illustrative purpose only). 

\subsection{Companion stars}
Apparently the most direct method is a search for a companion star in pre-SN images. This requires deep high-spatial resolution images for very nearby SNe Ia, ideally ones taken with {\em the Hubble Space Telescope (HST)}. This is practically challenging, but fortunately there were two extremely nearby `normal' SNe Ia in this decade - SNe 2011fe in M101 and 2014J in M82. For SN 2011fe, an RG companion is rejected, and a MS companion with $\gsim 1 M_{\odot}$ is also rejected.\cite{li2011} For SN 2014J, an RG companion is rejected, as well as a part of He donor systems.\cite{kelly2014} These limits disfavor a part of the SD channels but not exclusively, i.e., either the MS companion or the spin-up/down scenario is left non-rejected. Interestingly, the story is different for SNe Iax. A blue point source has been detected in a pre-SN image of SN Iax 2012Z,\cite{mccully2014} which is consistent with a single He star. If this is confirmed to be a companion, it would point to a He donor path to the explosion. 

Another strong diagnostics is a search for a surviving companion star. This requires systems that were already exploded in the past, so SN remnants (SNRs) either in the Milky Way (MW) or the Large/Small Magellanic Cloud (LMC/SMC) are ideal targets thanks to their proximity. A case for Tycho SNR is still controversial -- while there is a strong candidate for a surviving MS companion,\cite{ruiz-lapuente2004} different researchers provide different arguments for or against the association.\cite{gonzalez2009,kerzendorf2013,bedin2014} A limit against a surviving RG companion has been placed on SN 1006.\cite{gonzalez2012} A strong case is found for LMC SNR 0509-67.5, for which the non-detection of stars near the kinematic center of the SNR has a deep limit that kills virtually all the SD channels,\cite{schaefer2012} except for the spin-up/down scenario in which the companion star already evolved to a He WD at the time of the explosion of the primary. As for surviving companions of extragalactic SNe, an interesting case was reported for SN Iax 2008ha (the faintest in the SN Iax class), where a `red' point source was detected.\cite{foley2014} However, it is not like a blue source in the pre-SN 2012Z image -- so, if the association is confirmed for both systems (2012Z and 2008ha), it will provide a new insight on the SN Iax progenitor system; this should require either multiple channels within SNe Iax or the change of the nature of the companion before and after the explosion of a primary WD, or the post-SN source might be associated with a compact remnant. 

There has also been an increasing attention as for how a non-degenerate companion star would manifest itself in SN observations. It has been predicted that the crush of the SN ejecta into the companion's envelope will deposit the thermal energy into the SN ejecta.\cite{marietta2000} This will create a blue excessive emission in the first few days, and the effect is more significant for a more extended companion star.\cite{kasen2010,kutsuna2015} The observational test had been generally negative for a non-degenerate companion star (but note that the observation is challenging), especially highlighted by SN 2011fe which was observed already within a day since the explosion with no positive signature of the SN-companion interaction.\cite{nugent2011,bloom2012,brown2012} However, recently there are (at least) two SNe Ia for which significant excessive emission especially in UV and a blue portion of optical wavelengths has been reported, and they are considered to be strong candidates for the interaction with a companion star. The first report was on a peculiar SN Ia iPTF14atg,\cite{cao2015} so it might appear (again) that the SD scenario might be associated with outliers. However, the second example of this excessive blue emission was a normal SN 2012cg,\cite{marion2016} and this could indicate that there might be at least a contribution from the SD path in normal SNe Ia. Another possible detection of such a feature is reported for normal SN Ia 2015F, which however favors a small radius ($\lsim 1 R_{\odot}$) if interpreted as a signature of a companion star.\cite{im2015} 

Another outcome of the SN-companion crush is the contamination of hydrogen into the SN ejecta. It has been suggested that hydrogen will lead to a substantial flux in H$_{\alpha}$ in a nebular phase (half an year and thereafter),\cite{mattila2005} or will show up as P$_{\beta}$ even earlier in the post-maximum phase (already at $\sim 1$ month after the explosion).\cite{maeda2014} Searches for H$_{\alpha}$ in late-time spectra of normal SNe result in non-detection, disfavoring the non-degenerate companion, especially an RG companion.\cite{leonard2007,lundqvist2013,maguire2016} No clear sign of P$_{\beta}$ has been reported, but a large sample of NIR spectra have not yet been searched for. Recently, a good NIR spectral series of SN 2014J has been presented, with the upper limit of $\sim 0.1M_{\odot}$ for the amount of stripped hydrogen from the search of P$_{\beta}$.\cite{sand2016}

\subsection{Circumstellar matter}
Another important diagnostics is the CSM environment -- the SD predicts a dirty CSM environment, through a wind from a companion star or an `optically thick wind' from the WD primary, or nova shells. The DD predicts clean environment. The CD predicts a dirty CSM. Analogous to core-collapse SNe, such a dense CSM is expected to lead to a detectable emission originated from the conversion of the kinematic energy to radiation energy through the crush. For a nominal mass loss rate from the SD system, it is expected that such emission is mostly seen as non-thermal emissions produced by shock-accelerated electrons, either as a synchrotron emission in radio or an inverse Compton in X-rays. So far, no radio nor X-ray emission has been reported for any SNe Ia. Especially deep limits have been placed for SNe 2011fe and 2014J.\cite{margutti2012,margutti2014,perez2014}

Despite the non-detection of radio or X-rays from SNe Ia, even more extreme cases have on the other hand been indicated. For the higher CSM density, the kinetic energy is thermalized and a large fraction of the released energy is emitted in the optical wavelength. This is a situation observed for SNe IIn. It has been recognized that among SNe IIn, there are indeed a class of objects in which the `SN component' is better described by an SN Ia rather than a core-collapse SN. They are believed to be an SN Ia explosion within a dense CSM and termed `Ia-CSM' (\S 2 and references therein). Interestingly, the SN component both in the light curve and spectra is in general classified as an 1991T-like SN Ia. 

An interesting question is whether SN 1991T-like SNe are generally linked to SNe Ia-CSM. The link has been strengthened by X-ray observations of Kepler SNR \cite{katsuda2015}. As compared to Tycho (a prototypical SN Ia) and SNR~0509-67.5 (1991T-like) for which a classification of the historical SNe is possible through an echo emission,\cite{krause2008,rest2008} an X-ray spectrum of Kepler SNR obtained by {\em Suzaku} shows stronger Fe lines than the other two SNRs whereas K-shell lines from intermediate-mass elements (IMEs: Si, S, Ar, and Ca) are similar to one another. This strongly  indicates that Kepler's SN was likely a bright Type Ia event. Fitting the spatially integrated spectra by a multi-component model, the masses of the ejecta and the swept-up ISM/CSM are estimated. The mass of Fe in the ejecta thus derived indicates that it was an SN 1991T-like event. 

The analysis of the CSM component in Kepler SNR leads to an interesting indication. The inferred mass ($\sim 0.3 M_{\odot}$) shows the similarity to what has been derived for SNe Ia-CSM. Combining this with  the current size of the SNR, the mass loss rate of $\sim 1.5 \times 10^{-5} M_{\odot}$ yr$^{-1}$ (for the wind velocity of $10$ km s$^{-1}$) is inferred, favoring the SD scenario. The derived ionization time scale suggests that the interaction between the SN ejecta and CSM started at $\gsim 200$ yrs after the explosion, unlike the typical time scale of weeks to months for SNe Ia-CSM. These observations suggest that potentially most, if not all, of SN 1991T-like SNe Ia are linked to SNe Ia-CSM, but only the systems having experienced a substantial mass loss just before the explosion are observed as SNe Ia-CSM.\cite{katsuda2015} 

A potentially strong addition here is a search for circumstellar (CS) echoes, i.e., either scattering (optical) or re-emission (NIR--IR) of SN photons on the CS dust grains, where the spatial scale of CSM to be studied (or equally the look-back time scale in the mass loss history) can be extended by an order of magnitude as compared to the diagnostics using the signature of hydrodynamical interaction between the SN ejecta and CSM. It has been shown that there is no signature of the CS echo for a sample of normal SNe Ia. For SNe Ia with a good coverage in NIR light curves like SN 2006X, the most extreme case of the SD scenario with an RG companion has been disfavored.\cite{maeda2015} However, we note that there is a possible echo in the IR reported for SN Iax 2014dt,\cite{fox2016} again seemingly pointing to the SD path for SN Iax objects. Further, there is a strong and long lasting NIR emission observed for one `Super Chandrasekhar SN Ia candidate' SN 2012dn, indicating that the over-luminous SNe Ia may be related to the SD scenario.\cite{yamanaka2016}

Another strong diagnostics is based on narrow absorption systems toward SNe Ia, being investigated with high-resolution spectra. Such features can be produced either by CSM or ISM, and deconvolving these two contributions to identify a CSM component is key. For this purpose, there are two approaches, one searching for the time variability and the other studying the Doppler shift statistically in a sample of SNe Ia. Following the discovery of time variable Na I D lines toward SN 2006X,\cite{patat2007} there are a few examples showing the time variability.\cite{simon2009} While the fraction of SNe Ia showing the time variability at the level of what was observed for SN 2006X is limited to less than $\sim 10$\%,\cite{blondin2009} at least $18$\% of SNe Ia show the variability at a lower but detectable level. \cite{sternberg2014} An extreme case is observed for PTF 11kx, which showed variable narrow absorption lines of Na I D and Ca II, and later showing the emergence of the emission components in Ca II and H$\alpha$.\cite{dilday2012} This is most naturally explained by a scenario that the CSM initially showing the absorption features was later crushed by the SN ejecta. 

While the fraction of SNe Ia showing the variable absorption lines is small, it would not reject the possibility that (at least) a part of the non-variable absorption lines could also be associated with CSM. Indeed, in the sample of SNe Ia showing the non-variable Na I D systems, there is a significant imbalance between the blueshifted and redshifted components, with the systems showing the blueshifted absorption lines are more frequently observed.\cite{sternberg2011} This indicates that a part of the narrow absorption systems are associated with an outflow, likely in CSM around SNe. We note that it is challenging to identify the non-variable absorption systems toward individual SNe either as CSM or ISM beyond statistics, which is a missing link to connecting these two approaches. In this aspect, the first robust argument has been presented for SN 2014J, where all the Na I D absorption systems are associated with ISM.\cite{maeda2016} While it is necessary to perform  similar analysis for at least a few other SNe before being able to connect this finding to the statistical study, it is interesting to note that SN 2014J shows very strong and complex blueshifted Na I D absorption components. 

\subsection{Modes of explosions}
Different explosion mechanisms should lead to different appearance of SNe Ia. So far the `best' model to `fit' to the observational light curves and spectral sequence/evolution, in the authors' possibly biased view, is the delayed detonation model -- especially those simulated under a spherically symmetric approximation. In this model, the `Phillips relation' could be explained by a different timing at which the deflagration is turned into the detonation. Within the 1D context this may be controlled by the transition density,\cite{iwamoto1999}  but one still has to clarify why different SNe have different condition for the transition. Another possible source of the diversity includes the metallicity and the fraction of carbon in the progenitor WD.\cite{umeda1999} The phenomenological 1D pure deflagration model, W7 model,\cite{nomoto1984} does equally a good job in reproducing the observed light curves and spectra.\cite{branch1985} 

Recent development in the field now allows the direct comparison between the simulated light curves and spectra based on multi-dimensional models and the observed properties.\cite{roepke2012} Overall, the multi-dimensional delayed-detonation models predict the light curve and spectral evolution reasonably good as compared to the observed ones. The fit is in general inferior to the fit by the (phenomenological) 1D models -- however it is not a fair comparison as the 1D models typically have a tuning parameter to fit to the observation. In terms of the light curves and spectra, there is one possible shortcoming in the multi-D delayed-detonation model sequence -- as long as the distribution for the initial sparks is used as an adjustable initial condition, the resulting diversity does not explain the Phillips relation.\cite{sim2013}

It has been suggested that the violent merger model and the double-detonation model could also explain the overall features in the light curves and spectra. For example, the violent merger model from a pair of $1.1 M_{\odot}$ WD and $0.9 M_{\odot}$ WD predicts a light curve and spectra compatible to those of SN 2011fe.\cite{roepke2012} A different combination of the WD masses would lead to a diversity of SN observables; for example, a merger of two similar masses, each $\sim 0.9 M_{\odot}$, is proposed for sub-luminous SN 1991bg-like SNe Ia.\cite{pakmor2010} One possible shortcoming of the violent merger scenario is an extremely asymmetric structure of the ejecta, which would lead to the diversity as a function of the viewing direction much beyond the observed diversity in the light curves and spectral features, as well as a high degree of polarization signals.\cite{bulla2016} Also, the violent merger model requires the systems in which the total mass exceeds $\sim 1.6 M_{\odot}$.\cite{sato2015} The expected number of such events is probably not sufficient to account for the main population of SNe Ia.\cite{ruiter2011,ablimit2016}

We note that the violent merger scenario is a sub-Chandrasekhar model in terms of nucleosynthesis point of view, as the density of the primary WD is a main function to determine the characteristic nucleosynthesis products. Another variant of the sub-Chandrasekhar model is the double-detonation model. The double detonation model could also do a good job in reproducing the observed light curves and spectra, while the investigation here has been mostly restricted to one-dimensional models. The similar configuration to the double detonation is expected for a variant of the violent merger with a He WD companion (or a CO WD with an He envelope), with the possible difference that the merger scenario would lead to more asymmetric SN ejecta structure. An advantage of the sub-Chandrasekhar models is that it could explain the Phillips relation as a function of the mass of the primary WD.\cite{sim2010}

In sum, there are pros and cons for each explosion model, but overall it is difficult to distinguish different models from the light curves and spectra in the early phase. A powerful smoking gun is suggested to clarify the mass of the primary WD. This issue can be tackled by searching for the `deflagration ash' at high density realized only in a massive WD near the Chandrasekhar mass.\cite{maeda2010a} In the delayed detonation model of the Chandrasekhar mass WD, the deflagration flames first create neutron-rich Fe-peak elements (e.g., $^{58}$Ni) at high density regions, through electron capture reactions. This requires the fuel density of $\gsim 10^{9}$ g cm$^{-1}$, and thus this process is missing in the sub-Chandrasekhar WD model. Strictly speaking, $^{58}$Ni can also be produced at low-density regions which experience Nucleo Statistical Equilibrium (NSE) through the detonation wave, but in these region the neutron-rich Fe-peak elements {\em coexist} with $^{56}$Ni. This provides an interesting (and powerful) test -- the Chandrasekhar mass WD model provides the amount and distribution of stable Fe-peak elements different from the sub-Chandrasekhar mass WD models. 

From the `abundance tomography',\cite{mazzali2007} it has been suggested that stable Fe-peak elements are abundant in the central region while the amount of $^{56}$Ni is small there. Also, the amount of stable Fe peak elements has been estimated to be $\sim 0.2 M_{\odot}$, which is too large for the models without the electron capture reactions. 

Yet another, indeed generally model-independent, argument for the need for the electron capture has been suggested from the analysis of late-time spectra.\cite{maeda2010b} While deriving the masses of each elements from the fluxes of emission lines would suffer from uncertainties in the model for thermal conditions, the kinematics is basically free from the uncertainty. It has been shown that the kinematic distribution of $^{56}$Ni (as observed as [Fe II] and [Fe III] with relatively high excitation energies) is different from that of stable Fe-peak elements (as observed as [Ni II] or [Fe II] with relatively low excitation energies). Specifically, the group of forbidden lines associated with the `$^{56}$Ni-rich region' show no-offset in the Doppler shift, the other lines associated to the `stable Fe/Ni-rich region' have diversity in their observed central wavelengths,\cite{maeda2010b,maeda2010c,blondin2012,silverman2013b} from blueshift to redshift spanning from  $\sim -2,000$ to $2,000$ km s$^{-1}$. The straightforward interpretation is that the $^{56}$Ni is distributed in an approximately spherical way, while the distribution of the stable Fe peak elements has an offset of $\sim 2,000$ km s$^{-1}$, the configuration well in line with the 2D off-set ignition model of the delayed detonation of the Chandrasekhar mass WD.\cite{maeda2010c}

Thanks to the low velocity, SNe Iax offer a wealth of information on the abundance content and distribution. For example, individual Co lines are clearly detected in the NIR,\cite{stritzinger2014} while in normal SNe Ia these are severally blended. This confirms the nature of SNe Iax as a thermonuclear explosion. Also, the [Ni II] is strong in late phases for many SNe Iax, while there is a large diversity.\cite{foley2013,stritzinger2014,stritzinger2015,yamanaka2015,foley2016} This suggests that the explosion should involve the high density environment (see above), but the condition in the thermonuclear ignition may be diverse. 

Along the same line, deriving the mass of the stable Ni content in MW or LMC/SMC SNRs would be a powerful diagnostics as well. Indeed, a Ni line has been identified in X-ray spectra of a few SNRs.\cite{yamaguchi2015} The observed Ni line shows a diversity in the flux, which may indicate a diversity in the production mechanism of stable Ni but the effect of different thermal conditions should also be taken into account. In any case, the detection would suggest that the electron capture reactions as a characteristic feature of the Chandrasekhar WD model could be a generic feature of SNe Ia. Especially, SNR 3C397 shows a very strong Ni line in its X-ray spectrum, and the derived mass of stable Ni clearly requires the electron capture reactions to have operated in the SN Ia leading to this SNR.\cite{yamaguchi2015}

The detection of $\gamma$-rays from $^{56}$Ni/Co/Fe decay chain is also worth mentioning. This is very basic prediction from the thermonuclear explosion of SNe Ia,\cite{clayton1969,milne2004,maeda2012} but the solid detection was only possible just recently for one SN -- SN 2014J thanks to its proximity. The 847 keV decay line was clearly detected by {\em INTEGRAL},\cite{churazov2014} confirming the basic nature of SNe Ia as a thermonuclear explosion beyond doubt. The Compton continuum was also detected, as was confirmed by the independent telescope ({\em Suzaku}) at $\sim 2\sigma$.\cite{terada2016} A combination of a line flux and a continuum flux has been suggested to be key diagnostics of the mass of the exploding WD,\cite{sim2008,summa2013} but the difficulty to correctly calibrate the flux of the continuum did not allow the discrimination between the Chandrasekhar and Sub-Chandrasekhar models, unfortunately, for SN 2014J. Another diagnostics is the early phase emission in the first month, when only models with a large amount of $^{56}$Ni near the surface are able to produce detectable emission. There is a report on the detection of the 158 keV line at $\sim 20$ days since the explosion,\cite{diehl2014} indicating that a model with a large amount of $^{56}$Ni near the surface is favored. Interestingly, the kinematics suggests that this cannot be distributed in a spherically symmetric manner to reproduce the observed Doppler shift, pointing either a ring-like\cite{diehl2014} or blobby structure.\cite{isern2016} In the future, a new technology of $\gamma$-ray detectors, e.g., an electron-tracking Compton camera, may hopefully lead to detection of radioactive $\gamma$-rays from a number of nearby SNe Ia.\cite{tanimori2015}

\begin{table}
\tbl{Progenitor Scenarios and Observational Constraints$^{\text a}$.}
{\begin{tabular}{@{}cccccccc@{}}
\toprule
 & 
\multicolumn{2}{c|}{Delayed Detonation} & 
\multicolumn{2}{c|}{Failed Deflagration}& 
Double Detonation&
\multicolumn{1}{c|}{Violent He}& 
Violent C\\
Observations   &
SD$^{\text b}$ &
\multicolumn{1}{c|}{DD} &
SD &
\multicolumn{1}{c|}{DD} &
SD &
\multicolumn{1}{c|}{DD} &
DD \\\colrule
 & & & \bf{Normal SNe Ia} & & & & \\
{\bf Companion} & & & & & & & \\
Direct & No & Yes &  No & Yes &   Maybe & Yes &   Yes \\
Early Emission & Maybe & Yes & Maybe & Yes & Maybe & Yes & Yes \\
Hydrogen & Maybe & Yes & Maybe & Yes & Yes & Yes & Yes \\
{\bf CSM} & & & & & & & \\
Radio/X & No & Yes & No & Yes & No & Yes & Yes \\
Echo & Maybe & Yes & Maybe & Yes & Yes & Yes & Yes \\
Abs. Sytems & Yes & Maybe & Yes & Maybe & Yes & Maybe & Maybe \\
{\bf Explosion} & & & & & & & \\
Spec./LC/pol. & Yes & Yes & No & No & Maybe & Maybe & Maybe \\
Nucleosynthesis & Yes & Yes & No & No & No & No & No \\
$\gamma$-ray (2014J) & Maybe & Maybe & ... & ... & Yes & Yes & Maybe
\\\colrule
 & & & \bf{SNe Iax} & & & & \\
{\bf Companion} & & & & & & & \\
Direct & Maybe & No &  Maybe & No &  Yes & No &   No \\
{\bf CSM} & & & & & & & \\
Echo & Yes & No & Yes & No & Yes & No & No \\
{\bf Explosion} & & & & & & & \\
Spec./LC/pol. & No & No & Yes & Yes & No & No & No \\
Nucleosynthesis & Yes & Yes & Yes & Yes & No & No & No \\\colrule
 & & & \bf{Over-Luminous} & & & & \\
{\bf CSM} & & & & & & & \\
Echo & Yes & No & Yes & No & Maybe & No & No \\
{\bf Explosion} & & & & & & & \\
Spec./LC/pol. & Yes & Yes & No & No & Maybe & Maybe & Maybe \\\colrule
 & & & \bf{Ia-CSM/1991T} & & & & \\
{\bf CSM} & & & & & & & \\
Interaction & Yes & No & Yes & No & Maybe & No & No \\
Abs. Sytems & Yes & No & Yes & No & Maybe & No & No \\
{\bf Explosion} & & & & & & & \\
Spec./LC/pol. & Yes & Yes & No & No & Yes & Yes & Yes \\\botrule
\end{tabular}}
\begin{tabnote}
$^{\text a}$ Note that the list might be biased by the authors' viewpoints. Sometimes different (seemingly conflicting) constraints are placed for SNe in the same category, so providing an answer as either `yes' or `no' is probably an oversimplification and may well be misleading in many situations. Also, note that there could be multiple paths for each class of SNe, so the list is based on an idea whether a given scenario would explain a major fraction of the SNe in each category. \\
$^{\text b}$ Note that the companion in this model category can be either an RG or a MS, or even a He WD. The constraints here are mostly for the RG companion, with less robust constraints on the MS. The case with the He WD companion at the time of the explosion should indeed be similar to the DD case. 
\end{tabnote}
\label{aba:tbl1}
\end{table}

\section{Concluding Remarks}\label{aba:summary}

There are variants of SN Ia progenitor and explosion models, and no single model seems to be perfect to satisfy all the observational constraints available to date. Given the large diversity of SNe Ia in various observables, it is indeed possible that SNe Ia are originating from multiple populations. Furthermore, there are always theoretical uncertainties to interpret the data, and indeed it is risky to rely on a single observational approach to discriminate different models. Keeping all of these caveats in mind, we hope that the summarizing list of the comparisons between the models and observations (Tab. 1) serves as a useful guide for future works. 

So far, assigning the progenitor paths to outliers seems like more straightforward than for normal SNe Ia, at least at a quick look. Possible detections of a companion star in a pre-SN image of SN Iax 2012Z and a possible companion or a left-over compact remnant in a post-SN image of SN Iax 2008ha indicate that these systems could be related to the SD channel. However, the difference in the natures of these detected objects is puzzling, i.e., a blue source for SN 2012Z and a red source in SN 2008ha, and it requires both theoretical investigation and further follow-up of these SNe. Also, we should stress that even if the blue source in the pre-SN image of SN 2012Z would be solidly identified as a He companion star, it should raise another important question regarding the link between the progenitor scenario and the explosion mode. Indeed, the most popular explosion model for SNe Iax, the failed deflagration model, is associated either with a RG/MS companion or no companion, in apparent contradiction to the blue source. 

For other classes of outliers, the inferred link is mostly based on the diagnostics through CSM. A strong argument for an association to the SD scenario exists for SNe Ia-CSM (but note that it is also consistent with the CD model). A huge mass loss before the explosion is required, pointing to an RG companion or an association to recurrent novae. It seems that a main fraction of SNe Ia-CSM are associated with SN 1991T-like SNe. A question then is if the opposite is a case -- namely, whether most (or all) SN 1991T-like SNe are associated with massive CSM. A recent work on Kepler SNR is very indicative, which suggests that SN 1991T-like SNe could be generally associated with massive CSM. This could be most naturally explained by a variant of the SD model in a system with a huge mass loss. 

As for over-luminous SNe (i.e., Super-Chandrasekhar candidates), recently there is a suggestion that at least one over-luminous SN Ia is associated with massive CSM, either an RG wind or nova shells. While there is still a single example, this could indicate that over-luminous SNe Ia might be related to the SD channel. The details of the explosion nature is however not yet clear. 

Finally, there are a number of observational constraints placed for normal SNe Ia, and the situation is less clear than for the outliers. Indeed, one has to take it in mind seriously that the apparent links between the outliers and the progenitor scenarios (as mentioned above) may merely reflect a small number of constraining observations for these outliers, and once the observational data are increasing the situation may well become more complicated than currently believed -- the same situation as normal SNe Ia have been encountered. Still, the advantages and disadvantages for different scenarios suggest (at least to the authors) that the delayed detonation is the best model for a mode of the explosion, while the information related to the environment (companion and CSM) generally prefers the DD scenario (or ones predicting the clean environment, e.g., a spin-up/down scenario in the SD model). So, if we simply rely on the score sheet for each scenario like Tab. 1, the explosion of Chandrasekhar mass WD through the secular evolution after the WD-WD merger could be a straightforward interpretation. However, this is still too early to make this strong conclusion, as many questions/challenges have been raised for this scenario from different directions as well, including whether the system can avoid the off-center deflagration (not to form an ONeMg WD), whether the number of such systems are sufficiently large to account for the main population of normal SNe Ia, and so on. Also, one should note that there is at least one normal SN Ia showing a possible signature of a non-degenerate companion, and at least a fraction of normal SNe Ia show  the association with detectable CSM. 

Another important issue in the progenitor scenario of normal SNe Ia is that relations to the outliers must be considered self-consistently. One might jump onto the conclusion that normal SNe Ia are from the DD systems and outliers are from the SD systems, but this is probably an oversimplification. For example, the existence of at least a single path of the SD systems may indeed suggest that there must be other variants of the SD systems as well. Also, it is not trivial if the SD and DD can coexist to contribute to different populations of SNe Ia, since the evolutionary processes toward the SD and DD are coupled -- for example, the existence of the SD channel suggests that the common envelope evolution must be avoided for those systems while the DD generally requires the common envelope evolution.

\section*{Acknowledgments}
The authors thank the organizers of the successful MG14 meeting, and all the participants to the BN4 session for stimulating discussion and help for editing this session summary. The authors also thank Robert Jantzen and Takashi Nagao for their help in preparing this manuscript. This work has been partly supported by JSPS KAKENHI (No. 23740141 and 26800100 for K. M.; No. 23340055 for Y. T.) from the Japanese Ministry of Education, Culture, Sports, Science and Technology (MEXT), and by World Premier International Research Center Initiative (WPI Initiative), MEXT, Japan.


\begin{thebibliography}{99}

\bibitem{phillips1999}
M.~M. Phillips, et al., {\em AJ} {\bf 118}, 1766 (1999).

\bibitem{permutter1999}
S. Permutter, et al., {\em ApJ} {\bf 517}, 565 (1999).

\bibitem{riess1998}
A.~G. Riess, et al., {\em AJ} {\bf 116}, 1009 (1998).

\bibitem{matteucci1986}
F. Matteucci, L. Greggio, {\em A\&A} {\bf 154}, 279 (1986).

\bibitem{kobayashi1998}
C. Kobayashi, et al., {\em ApJ} {\bf 503}, L155 (1998).

\bibitem{ruiter2009}
A.~J. Ruiter, K. Belczynski, C. Fryer, {\em ApJ} {\bf 699}, 2026 (2009).

\bibitem{whelan1973}
J. Whelan, I. Iben, Jr., {\em ApJ} {\bf 186}, 1007 (1973).

\bibitem{nomoto1982}
K. Nomoto, {\em ApJ} {\bf 253}, 798 (1992).

\bibitem{iben1984}
I. Iben, Jr., A.~V. Tutukov, {\em ApJ} {\bf 284}, 719 (1984).

\bibitem{webbink1984}
R.~F. Webbink, {\em ApJ} {\bf 277}, 355 (1984).

\bibitem{sparks1974}
W.~M. Sparks, T.~P. Stecher, {\em ApJ} {\bf 188}, 149 (1974).

\bibitem{soker2015}
N. Soker, {\em MNRAS} {\bf 450}, 1333 (2015).

\bibitem{hicken2009}
M. Hicken, et al., {\em ApJ} {\bf 700}, 331 (2009).

\bibitem{nugent1995}
P. Nugent, et al., {\em ApJ} {\bf 455}, L147 (1995).

\bibitem{filippenko1992a}
A.~V. Filippenko, et al., {\em ApJ} {\bf 384}, L15 (1992).

\bibitem{filippenko1992b}
A.~V. Filippenko, et al., {\em AJ} {\bf 3104}, 1543 (1992).

\bibitem{howell2006}
D.~A. Howell, et al., {\em Nature} {\bf 443}, 308 (2006).

\bibitem{hicken2007}
M. Hicken, et al., {\em ApJ} {\bf 669}, L17 (2007).

\bibitem{yamanaka2009}
M. Yamanaka, et al., {\em ApJ} {\bf 707}, L118 (2009).

\bibitem{scalzo2010}
R.~A. Scalzo, et al., {\em ApJ} {\bf 713}, 1073 (2010).

\bibitem{chakradhari2014}
N.~K. Chakradhari, et al., {\em MNRAS} {\bf 443}, 1663 (2014).

\bibitem{li2003}
W. Li, et al., {\em PASP} {\bf 115}, 453 (2003).

\bibitem{phillips2007}
M.~M. Phillips, et al., {\em PASP} {\bf 119}, 360 (2007).

\bibitem{sahu2008}
D.~K. Sahu, et al., {\em ApJ} {\bf 680}, 580 (2008).

\bibitem{foley2009}
R.~J. Foley, et al., {\em AJ} {\bf 138}, 376 (2009).

\bibitem{foley2013}
R.~J. Foley, et al., {\em ApJ} {\bf 767}, 57 (2013).

\bibitem{hamuy2003}
M. Hamuy, et al., {\em Nature} {\bf 424}, 651 (2003).

\bibitem{aldering2006}
G. Aldering, et al., {\em ApJ} {\bf 650}, 510 (2006).

\bibitem{taddia2012}
F. Taddia, et al., {\em A\&A} {\bf 545}, L7 (2012).

\bibitem{silverman2013a} 
J.~M. Silverman, et al., {\em ApJS} {\bf 207}, 3 (2013).

\bibitem{leloudas2015}
G. Leloudas, et al., {\em A\&A} {\bf 574}, 61 (2015).

\bibitem{wheeler2012}
J.~C. Wheeler, 2012, {\em ApJ} {\bf 758}, 123 (2012).

\bibitem{aizu1973}
K. Aizu, {\em Prog. Theor. Phys} {\bf 49}, 1184 (1973). 

\bibitem{hoshi1973}
R. Hoshi, {\em Prog. Theor. Phys} {\bf 49}, 776 (1973). 

\bibitem{ezuka1999}
H. Ezuka, M. Ishida, {\em ApJS} {\bf 120}, 277 (1999). 

\bibitem{terada2001}
Y. Terada, et al., {\em MNRAS} {\bf 328}, 112 (2001). 

\bibitem{ishida1997}
M. Ishida, et al., {\em MNRAS} {\bf 287}, 651 (1997). 

\bibitem{yuasa2010}
T. Yuasa, et al., {\em A\&A} {\bf 520}, 25 (2010). 

\bibitem{hayashi2014a}
T. Hayashi, M. Ishida, {\em MNRAS} {\bf 438}, 2267 (2014).

\bibitem{hayashi2014b}
T. Hayashi, M. Ishida, {\em MNRAS} {\bf 441}, 3718 (2014).

\bibitem{hachisu2000}
I. Hachisu, M. Kato, T. Kato, K. Matsumoto, {\em ApJ} {\bf 528}, 97 (2000).

\bibitem{hachisu2001}
I. Hachisu, M. Kato, {\em ApJ} {\bf 558}, 323 (2001).

\bibitem{schaefer2011}
B.~E. Schaefer, {\em ApJ} {\bf 742}, 112 (2011).

\bibitem{santander2015}
M. Santander-Garc\'ia, et al., {\em Nature} {\bf 519}, 63 (2015).

\bibitem{li1997}
X.~-D. Li, E.~P.~J. van den Heuvel, {\em A\&A} {\bf 322}, L9 (1997).

\bibitem{hachisu1999a}
I. Hachisu, M. Kato, K. Nomoto, {\em ApJ} {\bf  522}, 487 (1999).

\bibitem{hachisu1999b}
I. Hachisu, M. Kato, K. Nomoto, H. Umeda, {\em ApJ} {\bf 519}, 314 (1999).

\bibitem{nomoto1984}
K. Nomoto, F.~-K. Thielemann, K. Yokoi, 1984, {\em ApJ} {\bf 286}, 644 (1984).

\bibitem{hachisu1996}
I. Hachisu, M. Kato, K. Nomoto, {\em ApJ} {\bf 470}, 97 (1996).

\bibitem{arnett1969}
W.~D. Arnett, {\em Ap\&SS} {\bf 5}, 180 (1969).

\bibitem{woosley1986}
S.~E. Woosley, R.~E. Taam, T.~A. Weaver, {\em ApJ} {\bf 301}, 601 (1986).

\bibitem{timmes1992}
F.~X. Timmes, S.~E. Woosley, {\em ApJ} {\bf 396}, 649 (1992).

\bibitem{livne1993}
E. Livne, {\em ApJ} {\bf 406}, L17 (1993).

\bibitem{khokhlov1991}
A.~M. Khokhlov, {\em A\&A} {\bf 245}, 114 (1991).

\bibitem{arnett1994}
D. Arnett, E. Livne, {\em ApJ} {\bf 427}, 315 (1994).

\bibitem{hoeflich1995}
P. H\"oflich, A.~M. Khokhlov, J.~C. Wheeler, {\em ApJ} {\bf 444}, 831 (1995).

\bibitem{hoeflich1998}
P. H\"oflich, J.~C. Wheeler, F.~-K. Thielemann, {\em ApJ} {\bf 1998}, 495 (1998).

\bibitem{iwamoto1999}
K. Iwamoto, et al., {\em ApJS} {\bf 125}, 439 (1999).

\bibitem{woosley2004}
S.~E. Woosley, S. Wunsch, M. Kuhlen, {\em ApJ} {\bf 607}, 921 (2004).

\bibitem{kuhlen2006}
M. Kuhlen, S.~E. Woosley, G.~A. Glatzmaier, {\em ApJ} {\bf 640}, 407 (2006).

\bibitem{roepke2007b}
F.~K. R\"opke, S.~E. Woosley, W. Hillebrandt, {\em ApJ} {\bf 660}, 1344 (2007).

\bibitem{roepke2007a}
F.~K. R\"opke, et al., {\em ApJ} {\bf 668}, 1132 (2007).

\bibitem{kasen2009}
D. Kasen, F.~K. R\"opke, S.~E. Woosley, {\em Nature} {\bf 460}, 869 (2009).

\bibitem{maeda2010a}
K. Maeda, et al., {\em ApJ} {\bf 712}, 624 (2010).

\bibitem{seitenzahl2013}
I.~R. Seitenzahl, et al., {\em MNRAS} {\bf 429}, 1156 (2013).

\bibitem{kromer2013}
M. Kromer, et al., {\em MNRAS} {\bf 429}, 2287 (2013).

\bibitem{livne1990} 
E. Livne, A.~S. Glasner, {\em ApJ} {\bf 361}, 244 (1990).

\bibitem{woosley1994}
S.~E. Woosley, T.~A. Weaver, {\em ApJ} {\bf 423}, 371(1994).

\bibitem{shen2009} 
K.~J. Shen, L. Bildsten, {\em MNRAS} {\bf 699}, 1365 (2009).

\bibitem{fink2010}
M. Fink, et al., {\em A\&A} {\bf 514}, 53 (2010).

\bibitem{yoon2005}
S.~-C. Yoon, N. Langer, {\em A\&A} {\bf 435}, 967 (2005).

\bibitem{hachisu2012}
I. Hachisu, M. Kato, H. Saio, K. Nomoto, {\em ApJ} {\bf 744}, 69 (2012).

\bibitem{boshkayev2013}
K. Boshkayev, J.~A. Rueda, R. Ruffini, I. Siutsou, {\em ApJ} {\bf 762}, 117 (2013).

\bibitem{wang2014}
B. Wang, et al., {\em MNRAS} {\bf 445}, 2340 (2014).

\bibitem{benvenuto2015}
O.~G. Benvenuto, et al., {\em ApJ} {\bf 809}, L6 (2015).

\bibitem{saio1985}
H. Saio. K. Nomoto, {\em A\&A} {\bf 150}, L21 (1985).

\bibitem{nomoto1991}
K. Nomoto, Y. Kondo, {\em ApJ} {\bf 367}, L19 (1991).

\bibitem{dessart2006}
L. Dessart, et al., {\em ApJ} {\bf 644}, 1063 (2006).

\bibitem{yoon2007}
S.~-C. Yoon, Ph. Podsiadlowski, S. Rosswog, {\em MNRAS} {\bf 380}, 933 (2007).

\bibitem{pakmor2010}
R. Pakmor, et al., {\em Nature} {\bf 463}, 61 (2010).

\bibitem{pakmor2012}
R. Pakmor, et al., {\em ApJ} {\bf 747}, 10 (2012).

\bibitem{sato2015}
Y. Sato, et al., {\em ApJ} {\bf 807}, 105 (2015).

\bibitem{ruiter2011}
A.~J. Ruiter, et al., {\em MNRAS} {\bf 417}, 408 (2011).

\bibitem{ablimit2016}
I. Ablimit, K. Maeda, X.~-D. Li, {\em ApJ}, 826, 53 (2016).

\bibitem{kushnir2013}
D. Kushnir, et al., {\em ApJ} {\bf 778}, 37 (2013).

\bibitem{aznar2013}
G. Aznar-Sigu\'an, et al., {\em MNRAS} {\bf 434}, 2539 (2013).

\bibitem{tanikawa2015}
A. Tanikawa et al., {\em ApJ} {\bf 807}, 40 (2015).

\bibitem{roepke2012}
F.~K. R\"opke, et al., {\em ApJ} {\bf 750}, 19 (2012).

\bibitem{li2011}
W. Li, et al., {\em Nature} {\bf 480}, 348 (2011).

\bibitem{kelly2014}
P.~L. Kelly, et al., {\em ApJ} {\bf 790}, 3 (2014).

\bibitem{mccully2014}
C. McCully, et al., {\em Nature} {\bf 512}, 54 (2014).

\bibitem{ruiz-lapuente2004}
P. Ruiz-Lapuente, et al., {\em Nature} {\bf 431}, 1069 (2004).

\bibitem{gonzalez2009} 
J.~I. Gonz\'alez Hern\'andez, et al., {\em ApJ} {\bf 691}, 1 (2009).

\bibitem{kerzendorf2013}
W.~E. Kerzendorf, et al., {\em ApJ} {\bf 774}, 99 (2013).

\bibitem{bedin2014}
L.~R. Bedin, et al., {\em MNRAS} {\bf 439}, 354 (2014).

\bibitem{gonzalez2012} 
J.~I. Gonz\'alez Hern\'andez, et al., {\em Nature} {\bf 489}, 533 (2012).

\bibitem{schaefer2012}
B.~E. Schaefer, A. Pagnotta, {\em Nature} {\bf 481}, 164 (2012).

\bibitem{foley2014}
R.~J. Foley, et al., {\em MNRAS} {\bf 443}, 2887 (2014).

\bibitem{marietta2000}
E. Marietta, A. Burrows, B. Fryxell, {\em ApJS} {\bf 128}, 615 (2000).

\bibitem{kasen2010}
D. Kasen, {\em ApJ} {\bf 708}, 1025 (2010).

\bibitem{kutsuna2015}
M. Kutsuna, T. Shigeyama, {\em PASJ} {\bf 67}, 54 (2015).

\bibitem{nugent2011}
P.~E. Nugent, et al., {\em Nature} {\bf 480}, 344 (2011).

\bibitem{bloom2012}
J.~S. Bloom, et al., {\em ApJ} {\bf 744}, L17 (2012).

\bibitem{brown2012}
P.~J. Brown, et al., {\em ApJ} {\bf 753}, 22 (2012).

\bibitem{cao2015}
Yi. Cao, et al., {\em Nature} {\bf 521}, 328 (2015).

\bibitem{marion2016}
G.~H. Marion, et al., {\em ApJ} {\bf 820}, 92 (2016).

\bibitem{im2015}
M. Im, et al., {\em ApJS} {\bf 221}, 22 (2015). 

\bibitem{mattila2005}
S. Mattila, et al., {\em A\&A} {\bf 443}, 649 (2005).

\bibitem{maeda2014}
K. Maeda, M. Kutsuna, T. Shigeyama, {\em ApJ} {\bf 794}, 37 (2014).

\bibitem{leonard2007}
D.~C. Leonard, {\em ApJ} {\bf 670}, 1275 (2007).

\bibitem{lundqvist2013}
P. Lundqvist, et al., {\em MNRAS} {\bf 435}, 329 (2013).

\bibitem{maguire2016}
K. Maguire, et al., {\em MNRAS} {\bf 457}, 3254 (2016).

\bibitem{sand2016}
D.~J. Sand, et al., {\em ApJ} {\bf 822}, L16 (2016).

\bibitem{margutti2012} 
R. Margutti, et al., {\em ApJ} {\bf 751}, 134 (2012).

\bibitem{margutti2014}
R. Margutti, et al., {\em ApJ} {\bf 790}, 52 (2014).

\bibitem{perez2014}
M.~A. P\'erez-Torres, et al., {\em ApJ} {\bf 792}, 38 (2014).

\bibitem{katsuda2015}
S. Katsuda, et al., {\em ApJ} {\bf 808},49 (2015).

\bibitem{krause2008}
O. Krause, et al., {\em Nature} {\bf 456}, 617 (2008).

\bibitem{rest2008}
A. Rest, et al., {\em ApJ} {\bf 680}, 1137 (2008).

\bibitem{maeda2015}
K. Maeda, T. Nozawa, T. Nagao, K. Motohara, {\em MNRAS} {\bf 452}, 3281 (2015).

\bibitem{fox2016}
O.~D. Fox, et al., {\em ApJ} {\bf 816}, L13 (2016).

\bibitem{yamanaka2016}
M. Yamanaka, et al., {\em PASJ}, accepted (arXiv:1604.02800) (2016).

\bibitem{patat2007}
F. Patat, et al., {\em Science} {\bf 317}, 924 (2007).

\bibitem{simon2009}
J.~D. Simon, et al., {\em ApJ} {\bf 702}, 1157 (2009).

\bibitem{blondin2009}
S. Blondin, et al., {\em ApJ} {\bf 693}, 207 (2009).

\bibitem{sternberg2014}
A. Sternberg, et al., {\em MNRAS} {\bf 443}, 1849 (2014).

\bibitem{dilday2012}
B. Dilday, et al., {\em Nature} {\bf 337}, 942 (2012).

\bibitem{sternberg2011}
A. Sternberg, et al., {\em Science} {\bf 333}, 856 (2011).

\bibitem{maeda2016}
K. Maeda, et al., {\em ApJ} {\bf 816}, 57 (2016).

\bibitem{umeda1999}
H. Umeda, et al., {\em ApJ} {\bf 522}, 43 (1999).

\bibitem{branch1985}
D. Branch, et al., {\em ApJ} {\bf 294}, 619 (1985).

\bibitem{sim2013}
S.~A. Sim, et al., {\em MNRAS} {\bf 436}, 333 (2013).

\bibitem{bulla2016}
M. Bulla, et al., {\em MNRAS} {\bf 455}, 1060 (2016).

\bibitem{sim2010}
S.~A. Sim, et al., {\em ApJ} {\bf 714}, L52 (2010).

\bibitem{mazzali2007}
P.~A. Mazzali, et al., {\em Science} {\bf 315}, 825 (2007).

\bibitem{maeda2010b}
K. Maeda, et al., {\em ApJ} {\bf 708}, 1703 (2010).

\bibitem{maeda2010c}
K. Maeda, et al., {\em Nature} {\bf 466}, 82 (2010).

\bibitem{blondin2012}
S. Blondin, et al., {\em AJ} {\bf 143}, 126 (2012).

\bibitem{silverman2013b}
J.~M. Silverman, et al., {\em MNRAS} {\bf 430}, 1030 (2013).

\bibitem{stritzinger2014}
M.~D. Stritzinger, et al., {\em A\&A} {\bf 561}, A146 (2014).

\bibitem{stritzinger2015}
M.~D. Stritzinger, et al., {\em A\&A} {\bf 573}, A2 (2015).

\bibitem{yamanaka2015}
M. Yamanaka, et al., {\em ApJ} {\bf 806}, 191 (2015).

\bibitem{foley2016}
R.~J. Foley, et al., {\em MNRAS} {\bf 461}, 433 (2016).

\bibitem{yamaguchi2015}
H. Yamaguchi, et al., {\em ApJ} {\bf 801}, L31 (2015). 

\bibitem{clayton1969}
D.~D. Clayton, S.~A. Colgate, G.~J. Fishman, {\em ApJ} {\bf 155}, 75 (1969).

\bibitem{milne2004}
P.~A. Milne, et al., {\em ApJ} {\bf 613}, 1101 (2004).

\bibitem{maeda2012}
K. Maeda, et al., {\em ApJ} {\bf 760}, 54 (2012).

\bibitem{churazov2014}
E. Churazov, et al., {\em Nature} {\bf 512}, 406 (2014).

\bibitem{terada2016} 
Y. Terada, et al., {\em ApJ} {\bf 823}, 43 (2016).

\bibitem{sim2008}
S.~A. Sim, P.~A. Mazzali, {\em MNRAS} {\bf 385}, 1681 (2008).

\bibitem{summa2013} 
A. Summa, et al., {\em A\&A} {\bf 554}, 67 (2013).

\bibitem{diehl2014} 
R. Diehl, et al., {\em Science} {\bf 345}, 1162 (2014).

\bibitem{isern2016}
J. Isern, et al., {\em A\&A} {\bf 588}, 67 (2016).

\bibitem{tanimori2015}
T. Tanimori, et al., {\em ApJ} {\bf 810}, 28 (2015).

\end{thebibliography}

\end{document}